\documentclass{desyproc}

\begin{document}
%------------------------------------
\title{The status and prospects of the Q  \textrm{\&} A  experiment with some applications}

%for single authors the superscripts are optional
\author{{\slshape Hsien-Hao Mei$^1$, Wei-Tou Ni$^1$, and Sheng-Jui Chen$^2$, Sheau-shi Pan$^2$ (Q \textrm{\&} A Collaboration) }\\[1ex]
$^1$Center for Gravitation and Cosmology, Department of Physics, National Tsing Hua University, Hsinchu, Taiwan 30013, Republic of China \ \ {\it mei@phys.nthu.edu.tw, wtni@phys.nthu.edu.tw}\\
$^2$Center for Measurement Standards, Industrial Technology Research
Institute, Hsinchu, Taiwan 30013 Republic of China}

% if the proceedings are available online (e.g. at Indico)
% please enter the contribution ID or file_name below for the DOI
%\contribID{32}
\contribID{lindner\_axel}

% TO THE CONFERENCE EDITORS:
% please update the following information
% before sending the template to the authors
% \confID{800}  % if the conference is on Indico uncomment this line
\desyproc{DESY-PROC-2009-05}
\acronym{Patras 2009} % if you want the Acronym in the page footer uncomment this line
\doi  % if there is an online version we will registerDOIs

\maketitle

\begin{abstract}
Motivated to measure the QED vacuum birefringence and to detect
pseudoscalar-photon interaction, we started to build up the Q
\textrm{\&} A experiment (QED [Quantum Electrodynamics] and Axion
experiment) in 1994. In this talk, we first review our 3.5 m
Fabry-Perot interferometer together with our results of measuring
Cotton-Mouton effects of gases. We are upgrading our interferometer
to 7 m armlength with a new 1.8 m 2.3 T permanent magnet capable of
rotation up to 13 cycles per second. We will use 532 nm Nd:YAG laser
as light source with cavity finesse around 100,000, and aim at 10
nrad/Hz$^{1/2}$ optical sensitivity. With all these achieved and the
upgrading of vacuum, QED birefringence would be measured to 28$\%$
in about 50 days. Along the way, we should be able to improve on the
dichroism detection significantly.
\end{abstract}

\section{Introduction}

In 1991, when Tsubono from University of Tokyo visited our
Gravitation Laboratory in Tsing Hua University, we discussed the
technical development of ultra-high sensitive interferometers for
gravity-wave detection. During the last day before his departure, we
pondered about how we could apply these developed techniques for
fundamental physics and we discussed the possibility of doing the
interferometric QED (Quantum Electrodynamics) tests. After analyzing
the sensitivities, we believed that the QED birefringence would be
measurable \cite{1}.

After the call for EOI's (Expression of Interest) of using the
onsite SSC (Superconducting Super Collider) facilities in March,
1994 by DOE of USA, we submitted a joint EOI with a US team
\cite{2}. The topic of this EOI was chosen as one of the six topics
for project definition study proposals. We then submitted such a
proposal \cite{3} in June and finished the study at the end of
October, 1994 \cite{4}. The project definition review was
well-received. A five-year proposal \cite{5} was submitted to the
National Science Council of the Republic of China for the ROC part
of the funding simultaneously. This proposal was approved in
January, 1995 pending on the approval of the US proposal of the
collaboration. Partial funding was allocated for the first year.
However, due to lack of potential funding of the US counterpart,
this program of collaboration was halted.

Nevertheless, in 1994, we started to build the experimental facility
for the Q \textrm{\&} A experiment (QED and Axion experiment) [6-8]
acquiring two vacuum tanks of the laser-interferometric gravity-wave
detector type and working on the measurement of mirror birefringence
\cite{9}. Since 1991 we have worked on precision interferometry --
laser stabilization schemes, laser metrology and Fabry-Perot
interferometers. With these experiences, we started in 1994 to build
a 3.5m/7m prototype interferometer for measuring vacuum
birefringence and improving the sensitivity of axion search as part
of our continuing effort in precision interferometry. In June, 1994
in the Marcel Grossmann Meeting at Stanford, we met the PVLAS
people, exchanged a few ideas and encouraged each other. We learned
that PVLAS also started in the same year adapting their earlier
scheme proposed in 1979 \cite{10}.

In 2002, we finished the first phase of constructing the 3.5 m
prototype interferometer and made some Cotton-Mouton coefficient and
Verdet coefficient measurements \cite{11}. Starting 2002, we have
been in the second phase of Q \textrm{\&} A experiment. The results
of our second phase on dichroism and Cotton-Mouton effect (CME)
measurement have been reported in \cite{12} and \cite{13}. In
section 2 and section 3, we review our achieved optical sensitivity
and summarize our gaseous CME measurement results. We are starting
the 3rd phase of our Q \textrm{\&} A experiment extending the 3.5 m
interferometer to 7 m with upgrades.  These together with the goal
of this phase will be presented in section 4. Section 5 concludes
with discussion and outlook.

\vspace{-3 mm}
\section{Achieved optical sensitivity}

The schematic of the present setup of our second phase is shown in
Fig. 2 of reference \cite{12} and Fig. 1 of reference \cite{13}.
These references gave details of the experimental setup. Fig. \ref{Fig:Apparatus}
shows a picture of the experimental apparatus. Our 3.5 m prototype
interferometer is formed using a high-finesse Fabry-Perot
interferometer together with a high-precision ellipsometer. The two
high-reflectivity mirrors of the 3.5 m prototype interferometer are
suspended separately from two X-pendulum-double pendulum suspensions
mounted on two isolated tables fixed to ground using bellows inside
two vacuum chambers. The sub-systems are described in [14-16, 12].
Our results in this phase give ($-$0.2 $\pm$ 2.8) $\times$
10$^{-13}$ rad/pass with 18,700 passes through a 2.3 T 0.6 m long
magnet for vacuum dichroism measurement, and limit
pseudo-scalar-photon interaction and millicharged fermions
meaningfully \cite{12}.

\vspace{-3 mm}
\section{Measurement of gaseous Cotton-Mouton effects}

Upon passing through a medium with transverse magnetic field,
linearly polarized light becomes elliptically polarized. Cotton and
Mouton first investigated this in detail in 1905, and the phenomenon
is known as Cotton-Mouton effect. We use our Q \textrm{\&} A
apparatus to measure the CMEs at wavelength 1064 nm in nitrogen,
oxygen, carbon dioxide, argon, and krypton in a magnetic field B =
2.3 T at pressure P = 0.5-300 Torr and temperature T = 295-298 K.
Our measured results are compiled in Table \ref{Table:CM} \cite{13}. For the
Cotton-Mouton coefficient, we follow the convention of \cite{17} and
use the normalized Cotton-Mouton birefringence $\Delta n_u$ at P = 1
atm and B = 1 T. Our results agree with the PVLAS results
\cite{18,19} in the common cases (Kr, N$_2$, O$_2$) within 1.2
$\sigma$. For Ar and CO$_2$ at 1064 nm, our results are new.

\vspace{-3 mm}
\section{Upgrades}

We are currently upgrading our interferometer from 3.5 m armlength
to 7 m armlength in the 3rd phase.  We have installed a new 1.8 m
2.3 T permanent magnet capable of rotation up to 13 cycles per
second to enhance the physical effects. Figure \ref{Fig:NewSetUp} shows the
configuration with our new magnet. We are working with 532 nm Nd:YAG
laser as light source with cavity finesse around 100,000, and aim at
10 nrad/Hz$^{1/2}$ optical sensitivity. {\it With all these achieved
and the upgrading of vacuum, QED birefringence would be measured to
28 $\%$ in about 50 days}. Along the way, we should be able to
improve on the dichroism detection significantly.  To enhance the
physical effects further, another 1.8 m magnet will be added in the
future. \setlength{\unitlength}{1 mm}
% %%%%%%%%%%%%%%Fig. 1 & 2
%%%%%fig1 (width height)(8,6)cm  fig1 (width height)(4.5,6)cm
%%%%%fig1,2 (width height)(12.5,6)cm

%\begin{figure}[tbph]
%    \centering
% {\includegraphics[bb=0 0 334 171,scale=0.35]{fig1,2.eps}}
%    \begin{minipage}[b]{14cm}
%    \parbox[t]{14cm}
%       {
%     {\vspace{0.1cm}\hspace*{1.2cm}\footnotesize Figure 1: A picture of experimental
%     apparatus.
%     \hspace*{0.9cm}Figure 2:~~A picture of new setup.}}
%     \end{minipage}
%\end{figure}

\begin{minipage}{14cm}
\begin{minipage}[h]{9cm}
\psfig{file=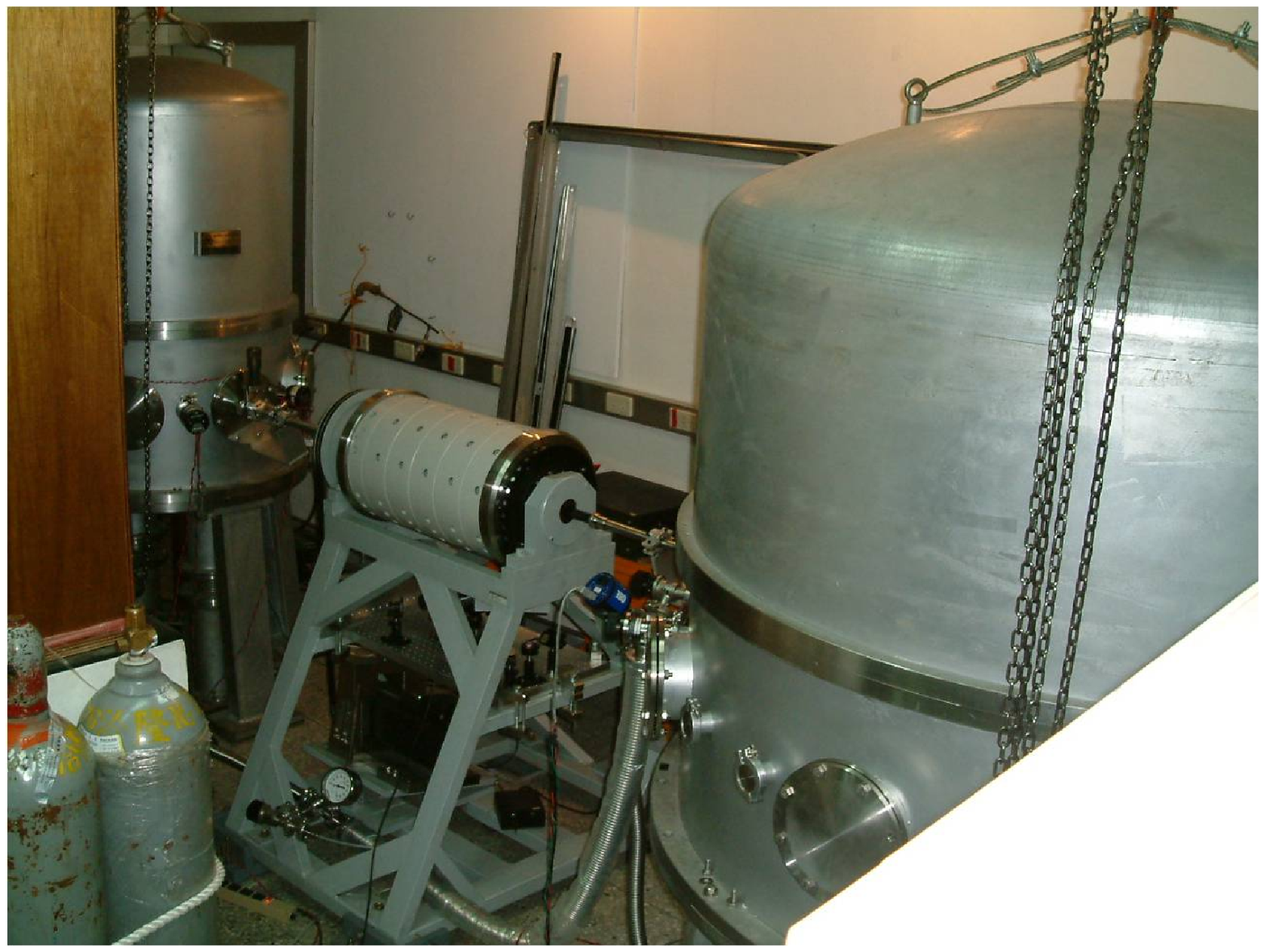,scale=0.525}
\makeatletter\def\@captype{figure}\makeatother\caption{A picture of experimental apparatus.}\label{Fig:Apparatus}
\end{minipage}
\begin{minipage}[h]{5cm}
\psfig{file=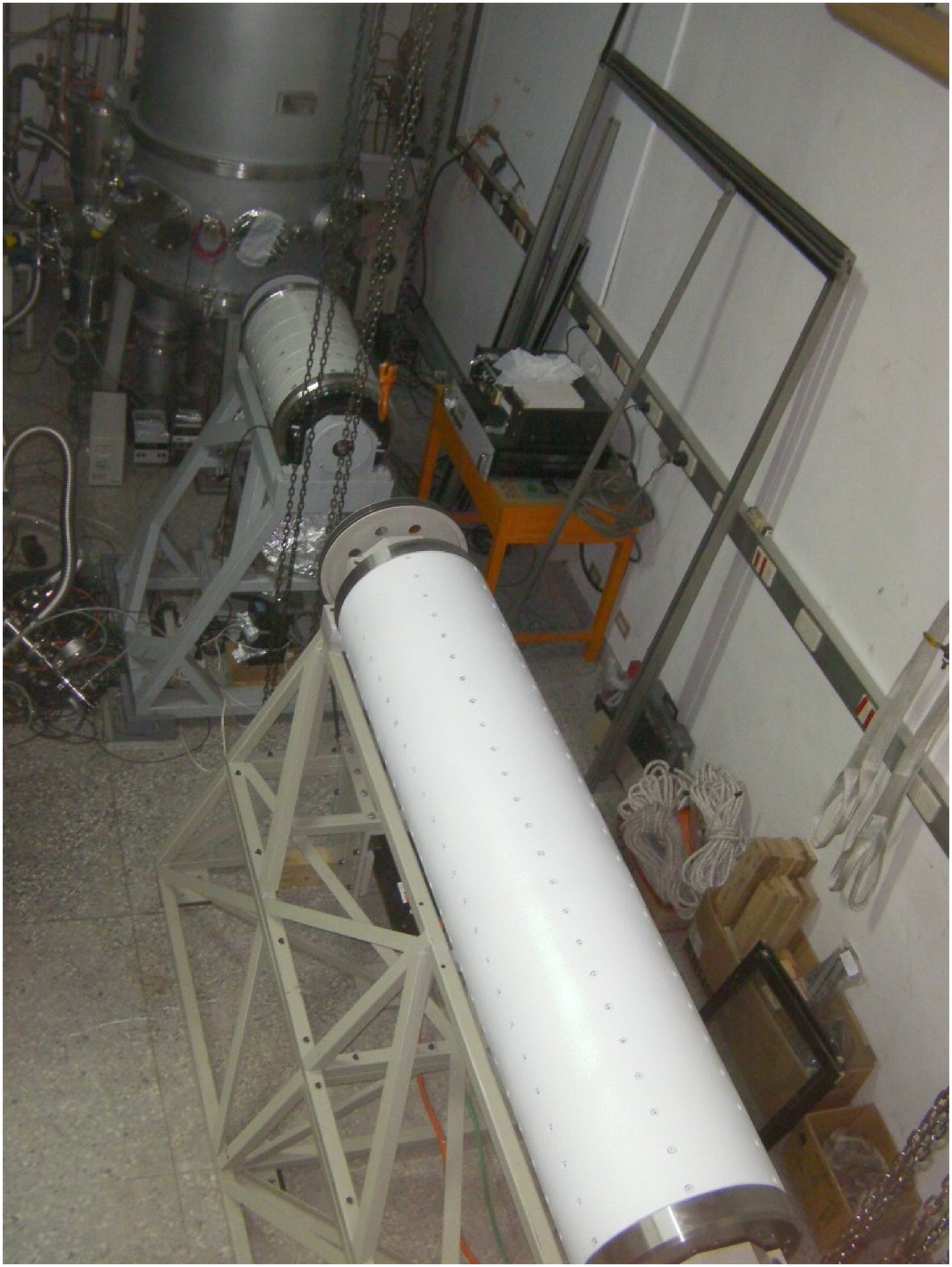,scale=0.187}
\makeatletter\def\@captype{figure}\makeatother\caption{A picture of new setup.}\label{Fig:NewSetUp}
\end{minipage}
\end{minipage}

\begin{table}[h]
\centering{\begin{tabular}{l} \hline\hline
 \\[-8pt]
Gas \ \ \ \ Normalized Cotton-Mouton birefringence  \\
 \ \ \ \ \ \ \ \ \ \ \ \ \ \ \ $\Delta n_u$ at P = 1 atm and B = 1 T \\[1pt]
\hline \\[-8pt]
N$_2$ \ \ \ \ \ ($-$ 2.02 $\pm$  0.16$^\S \pm$    0.08$^\P$) $\times$  10$^{-13}$ \\
O$_2$ \ \ \ \ \ ( $-$1.79  $\pm$ 0.34$^\S \pm$  0.08$^\P$)   $\times$ 10$^{-12}$ \\
CO$_2$ \ \ \   ($-$ 4.22 $\pm$  0.27$^\S \pm$ 0.16$^\P$) $\times$   10$^{-13}$ \\
 Ar  \ \ \ \ \ (4.31 $\pm$  0.34$^\S \pm$  0.17$^\P$)  $\times$  10$^{-15}$  \\
 Kr  \ \ \ \ \ (8.28 $\pm$  1.26$^\S \pm$  0.32$^\P$)   $\times$ 10$^{-15}$  \\[1pt]
\hline\hline $\S$: Statistical uncertainty\\
$\P$: Systematic uncertainty
\end{tabular}}
\caption{Measured Cotton-Mouton coefficients \cite{13}.}\label{Table:CM}
\end{table}

\section{Discussion and outlook}

We have heard a suite of motivations to search for
(pseudo)scalar-photon interactions and to measure QED birefringence
effect in this Patras 2009 workshop (See \cite{20} and other
articles in these proceedings; we refer the readers to various other
experiments to the proceedings also). For QED birefringence, the
next stage after detection is to measure the next-order effects
which include hadron and potential new physical contribution
\cite{8}. This would be possible by extending the interferometer
further with more rotatable permanent magnet. Many useful techniques
have been developed in the Gravitational Wave Detection Community.
We have advocated using relevant techniques \cite{1}. Recently,
there is a proposal to use the VIRGO facility \cite{21}. Further
progress in this experimental field is expected in the near
future.

\vspace{1 mm}
We thank the National Science Council (NSC
96-2119-M-007-004, NSC 97-2112-M-007-002, NSC 97-2811-M-007-057, NSC
98-2112-M-007-009, and NSC 98-2811-M-007-033) for supporting the Q
\textrm{\&} A program.

% ****************************************************************************
% BIBLIOGRAPHY AREA
% ****************************************************************************

\begin{footnotesize}
% IF YOU DO NOT USE BIBTEX, USE THE FOLLOWING SAMPLE SCHEME FOR THE REFERENCES
% ----------------------------------------------------------------------------

% ----------------------------------------------------------------------------

% IF YOU USE BIBTEX,
% - DELETE THE TEXT BETWEEN THE TWO ABOVE DASHED LINES
% - UNCOMMENT THE NEXT TWO LINES AND REPLACE 'Name_Of_Your_BibFile'

%\bibliographystyle{unsrt}
%\bibliography{Name_Of_Your_BibFile}

\begin{thebibliography}{99}
%------- replace following references ;-)

 \bibitem{1}W.-T. Ni {\it et al}., Mod. Phys. Lett. \textbf{A6} 3671 (1991).
\bibitem{2} ``Light Retardation in a High Magnetic Field and
Search for Light Scalar/Pseudo-Scalar Particles Using
Ultra-Sensitive Interferometry," Joint EOI (Expression of Interest)
submitted to the National Science Council of the Republic of China
and the Department of Energy of the United States of Amenica (April,
1994).
\bibitem{3}``Definition Studies for a proposal to Measure the Velocity of
Light in a Magnetic Field", proposal submitted to DOE of USA (June,
1994); ``Experimental Search for Spin-Coupling Interactions and
Light Scalar Sr: Pseudo-Scalar Particles," Prototype Study Proposal
submitted to the National Science Council of the Republic of China
(June, 1994).

\bibitem{4} U.S. Department of Energy, R-912, Light Retardation-Absorption
and Axion (LIRA) Experiment -- A Project Definition Study for
On-Site Use of the Superconducting Super Collider Assets and
Facilities, Final Report, Revision 1 (November 1994).
\bibitem{5} Test of Quantum Electrodynamical Birefringence and Search for Light
Scalar/Pseudoscalar Particles, A Five-Year Proposal submitted to
National Science Council (September, 1994).
 \bibitem{6} W.-T. Ni {\it et al}., in Proceedings of the Sixth Marcel Grossmann Meeting on General
Relativity, Stanford, California, 1994 (World Scientific, Singapore,
1996).
 \bibitem{7}W.-T. Ni, Chin. J. Phys. {\bf 34} 962 (1996).
 \bibitem{8} W.-T. Ni, Frontier Test of QED and Physics of the Vacuum, ed. E.
Zavattini, {\it et al}. (Sofia: Heron Press, 1998) p. 83;
 \bibitem{9} H.-W. Cheng {\it et al}, in International Workshop on Gravitation and Cosmology, December 14-17 (Tsing Hua Univ., Hsinchu, 1995), p.190.
 \bibitem{10} E. Iacopini and E. Zavattini, Phys. Lett. {\bf 85B} 151 (1971).
 \bibitem{11} J.-S. Wu, S.-J. Chen and W.-T. Ni, Class. Quantum Grav. {\bf 21} S1259 (2004).
 \bibitem{12} S.-J. Chen, H.-H. Mei and W.-T. Ni (Q \textrm{\&} A Collaboration), Mod. Phys. Lett. {\bf A22} 2815 (2007) [arXiv:hep-ex/0611050].
 \bibitem{13} H.-H. Mei {\it et al.}, Chem. Phys. Lett. {\bf 471} 216 (2009).
 \bibitem{14} S.-J. Chen, S.-H. Mei and W.-T. Ni, Improving ellipticity detection sensitivity for the Q \textrm{\&} A vacuum
birefringence experiment, hep-ex/0308071 (2003).
 \bibitem{15} S.-J. Chen, H.-H. Mei and W.-T. Ni, J. of Phys.: Conf. Series  {\bf 32} 244 (2006)
 \bibitem{16} H.-H. Mei, S.-J. Chen and W.-T. Ni, J. of Phys.: Conf. Series {\bf 32} 236 (2006).
 \bibitem{17} C. Rizzo, A. Rizzo, D. M. Bishop, Int. Rev. Phys. Chem. {\bf 16} 81 (1997).
 \bibitem{18} F. Brandi {\it et al.}, J. Opt. Soc. Am. {\bf B15} 1278 (1998).
 \bibitem{19} M. Bregant {\it et al}., Chem. Phys. Lett. {\bf 392} 276 (2004).
 \bibitem{20} W.-T. Ni, Constraints on pseudoscalar-photon interaction from CMB
polarization observation, these proceedings.
\bibitem{21} G. Zavattini and E. Calloni, Eur. Phys. J. {\bf C62} 459 (2009).



\end{thebibliography}
% example of Name_Of_Your_BibFile.bib
% @Article{Turcato:2006ch,
%      author    = "Turcato, M.",
%  collaboration = "ZEUS and H1",
%      title     = "Lepton flavour violation and charmonium physics at HERA",
%      journal   = "Nucl. Phys. Proc. Suppl.",
%      volume    = "162",
%      year      = "2006",
%      pages     = "283-287",
%      SLACcitation  = "%%CITATION = NUPHZ,162,283;%%"
% }
%
% @Unpublished{Gogitidze:2007du,
%      author    = "Gogitidze, N.",
%  collaboration = "H1",
%      title     = "Prompt photons and particle momentum distributions at
%                   HERA",
%      year      = "2007",
%      note    = "hep-ex/0701033",
%      SLACcitation  = "%%CITATION = HEP-EX 0701033;%%"
% }

\end{footnotesize}

% ****************************************************************************
% END OF BIBLIOGRAPHY AREA
% ****************************************************************************

\end{document}